**On properly integrating the electronic Raman and optical infra-red spectra of HTSC cuprate materials.**

**John A. Wilson**

H. H. Wills Physics Laboratory
University of Bristol
Bristol BS8 1TL U.K.

**Abstract**

New electronic Raman and I.R. spectroscopy results from optimally and overdoped high temperature superconducting (HTSC) cuprate systems are interpreted in terms of the negative-$U$, boson-fermion crossover model. Distinction is made between those features which follow the condensate gap, $2\Delta(p)$, and those that are set by the local-pair binding energy, $U(p)$. The critical role of doping level $p_c = 0.185$ is highlighted in conjunction with the matter of developing quasiparticle incoherence, making connection here with recent transport and related results. $E // c$ IR results in magnetic fields // and ⊥ $c$ prove particularly illuminating. The general scheme developed continues to embrace all experimental data very satisfactorily.

e-mail: john.a.wilson @bris.ac.uk

## §1. Introduction to boson-fermion negative-*U* crossover modelling of HTSC cuprates.

It has been notoriously difficult to come to any consensus on the interpretation of the highly characteristic, low energy optical behaviour of the high temperature superconducting cuprates. It has long been evident here, as with all other of their properties, that in neither superconducting nor normal state do these materials behave as simple classical systems [1,2]. There is self-evidently far too much abnormality for any conventionally based explanation to succeed in embracing all the observations. Not only is it necessary to fold suitably into the analysis now fairly routine matters like their proximity to the Mott transition and to a band structural saddle-point crossover, to their structural instability and charge striping, to their spin and charge density wave tendencies, but above all it is necessary to recognize the unusual strong scattering characterizing normal and superconducting state alike. What is source to the displayed Marginal Fermi Liquid behaviour? It is very clear that properly tracking and understanding the course and causes of this novel scattering will carry one to the heart of the HTSC phenomenon [3]. Some have turned to phonon soft modes and others to spin fluctuations, each to considerable acclaim, but to the author it never has seemed that these phenomena are sufficiently unusual or extreme to bestow upon the cuprates their manifest uniqueness.

Now charge fluctuations in an inhomogeneous setting, such as afforded by the present mixed-valent HTSC materials, in particular in view of their proximity to the Mott-Anderson transition, are a different matter. Coulomb interactions cover an altogether different energy scale. Of course this of itself is a problem, potentially negating their relevance. How can charge fluctuations be entertained in a near-Mott insulator? Surely, if given positive-*U* Hubbard energies ~ 4 eV, one is obliged to project them out of the problem. Now this might well be the case were the HTSC systems homogeneous, but they are not, and furthermore it could be true were the charge not directly engaged in chemical bonding, but it is. Not only are the electronic system and the lattice irrevocably one, interdependent, but within that whole the various band and bond energies are able to respond individually and simultaneously, both up and down. Charge fluctuations have complex courses and often striking outcomes [4] and none the more so than when quantum shell closure is involved.

In the present case of the mixed-valent cuprates [5], the double-loading charge fluctuations in question are adequately represented by the intervalence, three-unit process

$$^{9}Cu_{II}^{0} + {}^{9}Cu_{II}^{0} + {}^{8}Cu_{III}^{0} \longrightarrow {}^{8}Cu_{II}^{1+} + {}^{8}Cu_{II}^{1+} + {}^{10}Cu_{III}^{2-} .$$

The double-loading occurs here into coordination units which nominally carry trivalent Madelung charging by virtue of local counter-cation substitution and/or excess interstitial oxygen content. Notably the double loading brings the local, shell-filling, configuration $p^6d^{10}$, and in its trivalent setting the $d^{10}$ shell becomes greatly stabilized. The latter effect is not just atomic physics but the consequence of the new electron count above upon the local cation/anion bonding interaction. The antibonding *p*/*d* hybridization which hitherto had held the chemical potential high at $d^8$, and even at $d^9$, now becomes terminated. The $d^{10}$ closed shell is freed to relax rapidly in energy to its new status, semi-core-like when uncharged though here much distended, whilst the oxygen *p* states, no longer constituting stabilized partner within the $dp\sigma$ bonding interaction, rise in energy up through the descending copper *d*-state set. The end product is that the local chemical potential at the doubly-loaded $Cu_{III}$ coordination unit, instead of being driven upwards, actually finds itself back in close resonance with $E_F$ as this extends across the surrounding delocalized, mixed-valent, Fermi sea. The local pair $e + e \rightarrow b$ fluctuational creation locates in this way in near-resonance with $E_F$ in the HTSC materials. The strong characteristic anisotropic scattering engendering Marginal Fermi Liquid behaviour is perceived as being of the above origin together with its ensuing partner process of *e*-on-*b* scattering, i.e. of quasiparticles on local pairs [5]. The latter scattering is small-angle in form and it becomes dominant partner around $p(T_c^{opt})$, at the stage when local pair definition and lifetime are most advanced.

To arrive at this picture calls for a Hubbard negative-*U* value of –1.5 eV per electron [6], roughly the size of the $d_{x2-y2}$ bandwidth. *U* holds here its standard definition with relation to single-loading – namely here the local trivalent condition $^9Cu_{III}^{1-}$ – of

$$U = (E_{n+2} - E_{n+1}) - (E_{n+1} - E_n) = (E_{n+2} + E_n) - 2E_{n+1} \quad \text{per pair.}$$

In the mixed-valent environment of the HTSC cuprates the global level of $E_F$ is principally dictated not by $^9Cu_{III}^{1-}$ but by the majority species $^9Cu_{II}^{0}$. This means that the negative-*U* state $^{10}Cu_{III}^{2-}$ will not be rendered degenerate with $E_F$ unless and until $|U|$ per electron comes to take on roughly the band-width energy ~ 1.5 eV (see figure 3 in [11]). Actual *binding* energies, cursive $\mathcal{U}$, for a local pair (per electron) thus are reached only as and when $|U|$ exceeds such a value.

This value of Hubbard *U* of –1.5 eV per electron for the HTSC materials is supported experimentally from a wide variety of laser pump-probe experimental results [7,8,9,10], discussed in [6,10,11,12] by the current author, and from Little and coworker's near-infra thermomodulation spectroscopy [13], discussed at length in [14]. As indicated above this value of $U \approx -1.5$ eV per electron is able to position the crucial state relating to resonance and successful Boson-Fermion crossover modelling at a small binding energy, cursive $\mathcal{U}$, amounting to ~ -50 meV ($\approx 2\Delta(0)$). The precise size of this small binding energy as a function both of *p* and of overall system covalency is on display, it is claimed by the present author in [15,16,17], in the scanning tunnelling spectroscopy experiments of Davis, Hanaguri and coworkers [18,19,20]. The detailed situation as a function of *p* I drew out originally, when working from the Seebeck results, in figure 1 of [21]. The latter figure, now slightly modified and reproduced as figure 1, presents how the local pair level stands relative to the *absolute* binding level of $E_F(p)$. The energies specified were extracted from figure 3 of [5a], itself deriving from the author's spectroscopic studies on TM materials summarized in fig. 7 of [22].

Careful distinction needs to be made in all this work between the evolution of $\mathcal{U}(p)$ and of $\Delta(p)$, the latter being the system-wide superconductive gap instigated for the entire system under the stimulus of the resonant pairs. ARPES, as with all other optical probes, records the convolution of the two features. In these HTSC cuprates local pair formation under *e*-*e* scattering arises most effectively from the saddle 'hot spots' to the zone corner (see figure 3 in [6]), whilst the bosons themselves, once created, scatter also most strongly from the heavy quasiparticles of the saddle regions. The anisotropy and form of these twin components to the resistive scattering as functions of $T$ and $p$ recently have been addressed again in [23] following Hussey *et al*'s detailed transport, AMRO and dHS studies of HTSC systems [24,25,26]. While the negative-$U$ state first drops below $E_F(p)$ in the vicinity of $p$ = 0.27, the optimal interaction between bosonic and fermionic subsystems does not occur until $p \approx 0.18$. This concentration of holes is optimal in the sense of providing a suitably localized milieu for the sustained number of superconducting pairs $n_s$ to acquire its highest value, as is made manifest in μSR/penetration depth studies [27]. At this point the condensation energy per carrier, as disclosed from electronic specific heat studies [28], comes to its maximum value. $T_c$ optimization, on the other hand, is not gained until the slightly lesser $p$ value of 0.16. As a result of rapidly growing quasiparticle incoherence below $p$ = 0.18, the greater locality and definition accruing to the local pair Madelung potential quickly become out-weighed by the declining coherent quasiparticle count. As the binding energy $U(p)$ goes on increasing into the underdoped 'pseudogap' regime it rapidly becomes decoupled from the fermionic system, and from the size of $\Delta(p)$ procured for the overall superconducting state. This is the so-called gapping dichotomy [29], in which $\Delta(p)$ tracks $T_c(p)$ and $n_s(p)$, dropping back with the reduction in $p$ count, whilst $|U(p)|$ continues to increase steadily as $p$ declines towards the Mott level and the onset of antiferromagnetism. $\Delta_o(p)$ is, note, ultimately prescribed through action in the nodal, 45$^o$ directions, while $U(p)$, by contrast, being bound only towards the outer parts of the Brillouin zone, stands as an axial, saddle, or antinodal entity.

With this appreciation of what is underway in the HTSC cuprates, we may set about clarifying those hitherto perplexing details of the electronic Raman and IR optical spectra.

**§2. Understanding the Raman spectra of HTSC materials.**

Raman spectra provide a register both of electron-phonon scattering and of carrier-carrier electronic scattering. Phonon scattering has revealed some useful information and has provided a valuable complement to neutron scattering in assessing the unusual soft mode behaviour evident in the HTSC phenomenon. I do not wish to go back into this interesting and illuminating area, discussed at length in [15] in the light of the neutron results from Chung *et al* [30] and Pintschovius *et al* [31], but shall dwell here on the continuum electronic scattering below 800 cm$^{-1}$ (100 meV). This too has been known from early times in the HTSC saga to be unusually structured [32,33], showing characteristically varied behaviour for the $B_{1g}$, $B_{2g}$ and $A_g$ symmetry spectra. There arise interesting temperature and frequency variations and spectral scalings that make it clear the scattering processes they record lie at the heart of the HTSC problem. The theory of electronic Raman spectroscopy is laid out in [2] and it suffices now to appreciate that for pseudo-tetragonal material the $B_{2g}$ (or $XY$) spectra will be most sensitive to inelastic scattering in the vicinity of the nodal direction in the HTSC cuprates, whilst the $B_{1g}$ (or $X'Y'$) spectra emphasize inelastic scattering around the antinodal regions. Hence we expect from our earlier discussions that the latter spectra primarily capture the formation and subsequent break up of local pairs, whilst the former will in the main monitor the pairing induced in the Fermi sea by the local pairs, i.e. they are going chiefly to yield information on $U(p)$ and on $\Delta(p)$, respectively. Because the induced superconductivity bears closer numerical relation to $T_c$, it is the $B_{2g}$ spectra which generally have become associated with superconductivity, leaving the $B_{1g}$ spectra in a rather anomalous role. However, as we shall see, the $B_{1g}$ spectra manifest all the desired characteristics to be associated with the local pairs and hence are intimately concerned with the sourcing of the global superconductivity.

Let us look at a list of attributes of the $B_{1g}$ electronic Raman spectrum below 1,000 $cm^{-1}$ forthcoming from [32,33] and from the illuminating new release by Munnikes [34]:

i) There is a slowly saturating background continuum of scattering across this entire energy range that does not change greatly with $T$ for $T > T_c$.

ii) Below $T_c$ appreciable spectral weight is lost from its low energy range below around 250 $cm^{-1}$ (30 meV) to become, in optimally and moderately overdoped material, peaked up around 300 to 600 $cm^{-1}$ (35 - 70 meV), details being dependent here upon doping and $T_c(p)$.

iii) In underdoped material the above peak's *energy* does not fall away with fall in $p$ and $T_c(p)$, but goes on rising in value as $p$ decreases.

iv) The peak's *amplitude* (over and above the normal state scattering) however rapidly fades away into the underdoped region.

v) In overdoped material, by contrast, the energy of the peak formed below $T_c$ diminishes with increasing $p$ value in a manner now roughly similar to $T_c(p)$ itself. By contrast, the amplitude of the peak scattering diverges as the peak energy recedes with $T_c(p)$ to zero around $p \sim 0.27$ [34]

vi) The peak energy in overdoped material drops off in fact somewhat more rapidly than does $T_c(p)$, destroying thereby the scaling of $\hbar\Omega_{pk}$ to $kT_c$ shown by optimally and slightly underdoped samples. (N.B. for the $B_{2g}$ spectra there exists a complete scaling between $\hbar\Omega_{pk}$ and $kT_c$)

vii) Above $T_c$ the background continuum spectral intensity, across all $\omega$, fits to photon scattering from a *single-boson* entity [33]. This is to be contrasted with the two-boson variance in the two-magnon scattering in evidence at higher energies for strongly underdoped, insulating material.

viii) Below $T_c$ there is not sufficient low energy spectral weight change in these $B_{1g}$ IR spectra to match a standard superconducting transition, it appearing as if a substantial number of quasi-particles remain unpaired [33].

Let us here make some further comments on the above same points:

i) Note there is no linear-in-$\omega$ intensity fall-off to low energies able to be attributed to anything other than the Bose factor; in particular there is no indication here of the Marginal Fermi Liquid behaviour that is evident in the $A_{1g}$ and $B_{2g}$ spectra.

ii) The often rather analogous changes found in the $B_{2g}$ spectra would uphold, though, some close tie-in of $B_{1g}$ events to the superconductivity.

iii) This growth in $B_{1g}$ peak energy is precisely the form of change noted earlier for $U(p)$ as the binding energy below $E_F$ of the local-pair state grows (see fig.1).

iv) The peak amplitude fades away as the withdrawing negative-*U* state rapidly decouples from the free quasiparticles.

v) Through the *over*doped system the value of $U(p)$ diminishes until the negative-*U* state becomes unbound. By this level of doping ($p \approx 0.27$) screening is very high, all pair lifetimes are now very short, and the photon scattering diverges in amplitude under pair dissolution.

vi) The strong coupling physics, which at lower doping pushed both $U(p)$ and $T_c(p)$ to high values, is replaced here by much weaker coupling behaviour, and the scaling ratio between $kT_c$ and $\hbar\Omega_{pk}$ drops back towards a *d*-wave mean-field-like value of $\hbar\Omega_{pk}/kT_c = 4.3$, rather than being as in optimally and underdoped material up around 5.5 to 6 in the $B_{2g}$ spectra, and even higher in $B_{1g}$ (up to 9).

vii) The assessed bosonic form to the thermal dependence of the scatterer responsible for the $B_{1g}$ spectra is well matched by negative-*U* local-pairs, in a way which the other "localized entities" mooted find much more problematic.

viii) As was indicated by the μSR work of Uemura and others [27], the cuprate superconducting condensation is not classical, it no longer simultaneously engaging all the carriers once below $T_c$. Moreover as the electronic specific heat work [28] makes clear, transient local pairing persists to far above $T_c$; HTSC is not driven in the BCS fashion.

ix) A final further point remains. Electronic events become impervious to standard temperatures at excitation energies beyond ~ 800 $cm^{-1}$ (or 100 meV). Above this energy there occurs a merger of the "normal" and superconducting state Raman spectra. This does not however mean that the local pairs vanish once $T$ > 300 K: in all cases we are left with the abnormal structureless background scattering clearly extending through to remarkably high energies in excess of 1 eV. This scattering sources the Marginal Fermi Liquid behaviour, widely in evidence elsewhere in transport and ordinary IR optical spectroscopy.

What now of the $B_{2g}$ and $A_{1g}$ electronic Raman spectra? How do they fit in?

Let us first turn to examine the $A_{1g}$ spectra and what is revealed of the underlying physics disclosed by this polarization in addressing the second new Raman paper from Masui *et al* [35]. This work concentrates again upon optimally and over-doped material and deals with the system (Y/Ca)$Ba_2Cu_3O_{7-\delta}$. The usage here in Y-123 of such Ca substitution allows distinction to be made between those results which are a direct consequence of the hole doping content *p* and those that are to be attributed to the active participation of the structural chains in shaping the crucial behaviour in the planes. First point to note is that the $A_{1g}$ peak energy and peak intensity, as for the $B_{1g}$ spectra, strongly register a discontinuity at $p = 0.185$. Although referred to throughout [35] as occurring at $p = 0.19$, all the data is sufficiently detailed to make identification here with $p = 0.185$ (for which I have my own reasons, laid out in [14,section 4.3]). The comparably sharp discontinuity witnessed at this same composition in the anisotropic resistive scattering work on LSCO from Hussey and coworkers [24] allows one to understand how the discontinuity, (first universally established at such $p$ via electronic specific heat analysis [28] and designated there as 'pseudogap onset') is to be associated with the termination of full quasiparticle coherence. That coherence loss has been presented in [5d,23] by the current author as issuing from negative-$U$ local pair creation and from the ensuing quasiparticle scattering by those resonant pairs. Coherence is, in consequence, first in jeopardy within the antinodal regions, but steadily spreads towards the nodal regions as $p$ is further reduced. The current Raman work suggests that in YBCO incoherence sets in slightly earlier perpendicular to the chains than occurs parallel to them, the result of inferior 3D coupling (see fig 4 in [35]). However by $p$ = 0.185 the peak behaviours for both $YY$ and $XX$ polarizations have become identical, and the $A_{1g}$ peak is there at its most pronounced. For higher $p$ the $A_{1g}$ and $B_{1g}$ peaks had been indistinguishable, but with $p$ = 0.185 something of great import arises, as figure 2 (based upon figure 2 of [35] but now augmented by lines to guide the eye) will reveal. At this juncture the two peaks separate. The $B_{1g}$ peak runs away up toward higher energies ~ 70 meV, whilst the $A_{1g}$ peak following a step increase in energy from 30 meV holds to a much lower trajectory which maximizes with $T_c$ at $p$ = 0.16 at an energy of about 40 meV. If we are to associate the $B_{1g}$ peak with individual local pairs and binding energy $U(p)$, then we shall associate the $A_{1g}$ peak with the superconducting condensate as engendered by the local pairs. The observed peak energy of 41 meV attained at optimal doping supports the earlier claim that the so-called neutron resonance peak, so prominent at $p^{opt}$ at this energy, is to be associated with a spin-flip excitation within the condensate from the spin-singlet to the spin-triplet condition. Such a process effects the immediate dissociation of any local pair involved as the two parallel spins no longer can quantum mechanically be accommodated within the same locality [15(§3)]. Towards lower $p$ and lower $T_c$ the $A_{1g}$ peak quite quickly fades away to become replaced as dominant feature by the $B_{2g}$ peak. The latter tracks the residue of induced $d$-wave superconductive coupling, associated now more and more with the nodal quasiparticles as interaction with the receding negative-$U$ state diminishes.

The maximization of the $A_{1g}$ peak intensity at $p$ = 0.185, right where the superconductive energy is greatest, emphasizes the driving role of the local pairs in securing HTSC behaviour. This sharp growth in amplitude of the $A_{1g}$ peak (relative to the $B_{1g}$ peak [35 fig.3b]) is a marker of the fact that the $A_{1g}$ symmetry susceptibility is much the more sensitive to electronic screening [36]. This point was made early on by Cardona and coworkers [37] when, via fine tuning the $A_{1g}$ *phonon* line across $T_c$ by employing the full gamut of rare-earth substitutions, it first was possible to pinpoint the superconducting gap energy. The presently observed (much larger) step adjustment in the electronic Raman peak to higher *energy* as one transits through the coherence limit at $p$ = 0.185 once again expresses the sensitivity of the $A_{1g}$ symmetry signal to decrease in electronic screening. It is evident from figure 2 that if this step up in $A_{1g}$ peak energy could be enhanced one might expect $T_c$(opt) to be enhanced. That, it would seem, is what in fact occurs as one manipulates the counter-ions involved in these systems: as for example in turning to the divalent Hg systems from the higher valent and lone-pair containing Tl-, Pb- and Bi-based systems. Similarly some substitution of oxygen by more ionic fluorine would appear to be a move in the same direction as borne out by the maximum $T_c$ reported to date coming from fluorine substituted Hg-1223 [38].

Finally a word about the $B_{2g}$ spectra. These spectra do not change greatly in amplitude with doping and their behaviour throughout reflects immediate association with the superconducting condensate. Unlike with $B_{1g}$, there exists full and universal scaling between $T_c$-normalized spectra over all $\omega$, $h\Omega_{pk}/kT_c$ holding throughout to the strong-coupling value of 5.5 to 6. Such a value is retained even as $T_c$ falls with $p$ to very low values. At *all* doping levels, as far as the mean-field, BCS-like, induced, more nodal component to the superconductivity is involved, HTSC stands the outcome of very strong and unusual coupling.

**§3. Understanding the infra-red optical spectra of HTSC systems.**

Study of the pair coupling in superconductors in traditional circumstances revolves around the function $\alpha^2.F(\Omega)$, the electron-phonon spectral density function. In the present more complex case one can generalize this prescription to the form $I^2.\chi(\Omega)$, relating to the 'active' bosonic spectral density, which may or may not involve retarded coupling. The interactions invoked do not directly manifest themselves in the optical (I.R.) spectra, but call for spectral inversion according to extended Drude, strong coupling-type procedures. Strong characteristic features which show up in the real and imaginary parts of the optical conductivity are related to strong features in $I^2.\chi(\Omega)$ that do not fall at quite the same energy.

The most advantageous mathematical fitting to extract the desired information from optical spectra is accomplished using the quantities the optical effective mass and the optical scattering rate. If one defines the optical self-energy $\Sigma^{op}(T,\omega)$ relating to the optical conductivity via the equation

$$\sigma(T,\omega) = \frac{\omega_p^2}{4\pi} \cdot \frac{i}{\omega - \Sigma^{op}(T,\omega)} ,$$

then the real part of $\Sigma^{op}(T,\omega)$ defines the relative optical effective mass ($m^{op}/m^{bare}$)

through $\qquad \omega[(m^{op}/m^b) - 1] = -2. \Sigma^{op}_1(T,\omega)$ ,

while the imaginary part of $\Sigma^{op}(T,\omega)$ defines the optical scattering rate through $\qquad 1/\tau^{op}(T,\omega) = -2. \Sigma^{op}_2(T,\omega)$ .

Note here the two-particle (*e-h*) optical scattering process will yield somewhat different information than does the single-particle scattering excitation experienced in photoemission.

Now what is the characteristic form taken by the HTSC IR spectra which for so long has perplexed researchers? The dominant feature characterizing both the real and imaginary parts of the experimental spectra is universally observed to sit at energies approaching *twice* that expected for $2\Delta_o$ as elicited from other types of measurement. On the real side, the self-energy spectrum takes the form of a quite sharp peak, followed by a much broader, lower intensity hump. On the imaginary side, one finds a step edge, followed by steadily rising losses. While in both cases the broader aspects here are retained to high temperature, the sharp features show up only below about $2T_c$.

The new IR paper from Yang, Hwang, Timusk and coworkers [39] has reported such $E\perp c$ spectra from good quality, slightly underdoped Hg-1201 crystals ($T_c$ = 91 K), and they have followed through the necessary spectral inversion, employing a maximum entropy routine, to access the underlying initiating spectral function $I^2.\chi(\Omega)$. While the experimental $m^{op}(T,\omega)$ and $1/\tau^{op}(T,\omega)$ plots display their characteristic changes up around 90 meV, it very tellingly emerges that the spectral coupling function responsible presents a very sharp peak back at 56 meV, this followed up by a weaker, much broader hump running out to high energy. This resonant feature in the coupling function at $\Omega_r$ = 56 meV note still stands considerably beyond the $2\Delta_o$ gap energy of 45 meV appertaining to the slightly underdoped sample. In addition one finds that energy $\Omega_r$ is effectively $T$-independent (see inset to fig.2 in [39]), while the appreciable thermal dependence of the resonance amplitude and width proves quite un-phonon-like, the feature becoming rapidly less pronounced as $T$ rises. All these observations support one, in light of what was said in §2, in making identification of the resonant feature in $I^2.\chi(\Omega)$ with $U(p)$, the negative-$U$ state binding level.

Yang *et al* next provide the results of a similar analysis for Hg-1223, employing earlier data, and reach the appropriately scaled values of 130 K, 58 meV and 72 meV for $T_c$, $\Delta_o$ and $\Omega_r$ respectively, the experimental spectral features in $1/\tau^{op}$ and $m^{op}$ now lying for Hg-1223 up near 110 meV. Taking a whole range of approximately optimally-doped materials the authors discover that $\Omega_r/kT_c \sim 6.3$ – though the detailed values for $\Omega_r(p)$ will have to track $U(p)$ if the above identification is correct. As is to be expected from §2, this number of 6.3, given the (sub)optimal doping, is somewhat greater than the analogous ratio of 5.5 relating to the neutron spin resonance peak. The latter figure, recall, refers not to the binding energy of the local pairs *per se* but to the overall superconducting condensate (see figure 2).

The analysis of such IR work has been advanced further in a very recent set of papers from van Heumen *et al* [40]. A direct attempt is there made to strip away all the more mundane thermal excitations for the bosons and fermions involved in the self-energy of the interaction process into a subsidiary multiplicative function so as to expose the fundamental frequency (and indeed ultimately temperature dependence) of the residual glue function. The latter is labelled now $\tilde{\Pi}(\omega,T)$ to express a more general regard for the actual interactions responsible for the witnessed self-energy changes. In particular, with regard to the resonance peak and the broader feature encountered above, an attempt is made to examine their individual characters. Are they both part of the pair-glueing operation, or does one in fact relate to pair-forming and the other perhaps to pair-breaking? It emerges that, while much of value can be recovered from the attempted customary strong-coupling treatment of the results, there clearly is not a full analogy here with what has been witnessed for phonon-mediated coupling, or is anticipated for spin fluctuation coupling.

What is undertaken in the more exploratory of the three above papers [40c] is a comparison of the outcome of these new experiments with what might be expected if (i) Marginal Fermi Liquid theory [3] were to prevail in the normal state, or (ii) the conditions implied by spin fluctuation theory held [41]. In truth neither of these looks appropriate because of the rather structureless forms of $\tilde{\Pi}$ implicit, and it is only after an additional low energy Lorenztian is rather artificially introduced that any real matching can be achieved to experiment. A direct evaluation of the $\tilde{\Pi}$ glue function from experiment discloses, by contrast, quite distinct interactions proceeding in the far- and the mid-IR. By extracting the coupling constant $\lambda$ using

$$\lambda = 2.\int^{\omega c}_0 \tilde{\Pi}(\omega)/\omega..d\omega ,$$

it is possible to obtain the following decomposition of $\lambda_{total}$ for Hg-1201 at 295 K [40c]:

$$\lambda_{total} = 1.85, \text{ with } \lambda_{FIR} = 0.7 \text{ and } \lambda_{MIR} = 1.2 .$$

Note because of the narrowness of the FIR resonance feature that, as far as coupling goes, it stands the junior partner here. It exhibits considerable broadening, however, at lower temperatures towards lower energies, and it is this which largely is responsible for the significant temperature dependence of $\lambda_{total}$. The latter mounts in Hg-1201 from 1.85 at room temperature, to 2.0 at 200 K, to 2.3 by just above $T_c$. Such values are very appreciably higher than the values for $\lambda \sim 0.3 - 0.5$ extracted from the nodal, low-energy dispersion kinks disclosed in ARPES work. Remember here though that optical spectroscopy is not *k*-sensitive and it will automatically now incorporate the antinodal response wherein the current superconductive coupling activity is pre-eminent.

How then does one account for the above apparent dual aspect to the coupling function in the negative-*U* scenario? If the resonance feature bears the signature of those local pairs rendered external to the condensate, what is the 200-300 meV MIR feature all about? Clearly the latter lies way above the standard range of harmonic phonon energies, and it exceeds, moreover, what is relevant for magnetic coupling in the HTSC cuprates [42]. Pointedly it resides precisely where the so-called 'waterfall' effect is found in ARPES work [43]

Let us reflect a little more closely upon the detailed nature of the local pairs in the present situation and upon how they are eliminated by optical excitation. The first stage is to unbind the negative-*U* pair relative to $E_F$; i.e. to supply the energy $U(p)$ of the resonant feature. The second stage is to shake off the lattice energy change associated with the local structural relaxation incurred with double loading, i.e. in the change from ${}^{9}Cu_{III}^{1-}$ to ${}^{10}Cu_{III}^{2-}$. What this involved was the Jahn-Teller distorted, *c*-axis elongated $d^9$ site in conversion to the spherical $d^{10}$ closed-shell condition. Additionally there occurred a distension of the closed-shell under the action of the (partially screened [4]) charge imbalance between the local nuclear charge and the site's acquired electron complement. These effects bring substantial changes to the lattice, long registered in measurements like EXAFS and PDF, or in isotopic $T_c$ shifts, etc.. More recently we have had the very striking structural results from Röhler [44] and from Gedik *et al* – the latter obtained using laser pump-probe crystallography [10;12]

Now in the decomposition of the local pair we still have not finished because the pair has yet to be broken electronically; i.e. the correlation energy has yet to be supplied to return the double-loading electron back into another coordination unit. In fact this is the major energy involved, and the pair disintegration stands accordingly not as a two-stage process but a three-stage process. It has been known for ten years or so that a further 1.5 eV per electron is demanded to see the pair revert back into two independent quasiparticles. Such information is forthcoming [11,14] from two different sources (i) laser pump-probe optical spectroscopy [7,8,9], and (ii) thermo-modulated, near-infra-red optical spectroscopy [13]. As was revealed from the latter work by Little and coworkers, despite the $1/\omega$ factor much diminishing the contribution to $\lambda$ issuing from such high energies, the NIR excitations bring a further augmentation ~ 0.35 to the $\lambda$ values actually operative.

The question that raises itself now is how, in the presence of such large $\lambda$ values, can $T_c$ manage only to be as high as it is, rather than breaking through to the long sought for target of 300 K. The answer must lie in the fact that these systems are not homogeneous nor indeed fully coherent. Fermi Liquid theory no longer holds here, with basic scalings such as the Wiedemann-Franz law becoming non-classical as $p$ progressively is reduced [45]. In particular, with regard to the superconductivity, any direct application of the equations within the customary Eliashberg formalism must prove inadequate.

The optical work itself provides an indication of where this breakdown in standard strong-coupling work is occurring and a glimpse as to what may yet be done to elevate $T_c$ further towards the great objective. By means of very careful FIR experimental and analytical investigation of the Drude limit, van der Marel and coworkers have shown [46] that the spectral weight changes which arise in the cuprates across $T_c$ are not in line with standard superconductive condensations. Normally on cooling below $T_c$ not only does the potential energy fall (stabilize), but the quasiparticles shed kinetic energy. However, by contrast, with the HTSC materials once below $p_c$ = 0.18 they exhibit what has been presented as an anomalous rise in K.E. on passing down through $T_c$, appearing in this to counter the fall in potential energy deriving from the superconductive condensation. Figure 3, based on fig.1 of [40c], displays the extracted spectral weight change results for a whole range of HTSC systems and dopings. The envelopes here sketched in should draw out the systematics of what is being disclosed. The upper envelope is a representation of what might be expected in a standard circumstance. The lower envelope indicates that the sharp deviation from such behaviour clearly is centred upon $p$ = 0.185. From the actual sign inversion in $\Delta W$ witnessed here and from the particular concentration involved, it would appear to the present author that interpretation of this development in terms of kinetic energy change is not the avenue to follow. It becomes more appropriate to set the discussion in terms of loss of coherence within the 'normal' state, and of the ensuing loss of full-time participation of the full complement of electrons within the superconducting condensate. We have witnessed from the specific heat work [28] how the condensation energy tumbles below this $p_c$, how the superfluid density as revealed by the μSR work [27] fades away, and how the zero-frequency transport scattering is transformed there [24]. The deficit in kinetic energy registered below $T_c$ must in part accordingly be due to the fact that it is not drawn from *all* the carriers in the customary way. Worries about what high-energy cut-off to employ in the spectral weight integral have on this count to stand without resolution. The unascertainable shortfall clearly is though in no way the driver of or indeed marker for the superconductivity itself, since $T_c$ shows no discontinuity at all at $p_c$. HTSC thrives on action at the edge of incoherence, where metallic screening is weak and local pair formation and interaction is best advanced.

### §4. The illumination of events provided by *E*//*c* IR spectra.

The above perception of the course of events is excellently endorsed in the very recent work from LaForge and colleagues [47]. These researchers have chosen to concentrate on the *c*-axis polarized spectra, rather then on the much more metallic and less strongly structured basal-plane response recorded in the more usual $E \perp c$ polarization. To obtain accurate data off the vertical faces of the crystals is of course much more demanding than securing $E \perp c$ basal plane spectra, but the *E*//*c* polarization proves particular sensitive to all *c*-axis coupling, and to coherence in general, whether for the normal or for the superconducting states. This high sensitivity may be tweaked further by the application of a magnetic field. Recognizing this, LaForge *et al* [47] have made a detailed spectral-weight study for YBCO-123 as a function of doping level, and contrasted the often quite slight but significant differences in spectral form which result when the reflectance experiments are carried out in magnetic fields of up to 8 tesla, including having *H*//*c* (as more normal) or $H \perp c$. That such changes and differences actually are registered under fields of this magnitude, far below $H_{c2}$, is marker of the fact that the primary deleterious effect of a magnetic field upon a superconductor is not as individual pair breaker, but as breaker of the phase coherence for the entire ensemble. The observation that the changes within the *H*//c and $H \perp c$ spectra may in places actually be rather large stresses the advantage of working in *E*//*c* polarization, for which all matters of coherence become ultra-sensitive when $p$ is at and just below $p_c$.

Let us now examine in some detail what is being observed in these *c*-axis spectra. Firstly it should be recalled that due to the much-decreased metallicity in this direction the reflectance IR no longer is dominated by the Drude response and the phonons now stand out clearly. Below $T_c$, when the spectrum is further strongly modified by formation of the condensate, the general transfer of spectral weight to zero frequency makes visible the *c*-axis Josephson plasma resonance (JPR) excitation for the condensate, as well as further collective excitations in the FIR. From the frequency of the JPR, strongly in evidence in reflection, it is possible to obtain a direct measure of $\rho_s$ itself for the sample via application of the equation $\omega_{JPR} = \sqrt{(\rho_s/\varepsilon_{\infty c})}$. $\rho_s$ is also derivable from $\omega.\sigma_2(\omega)$, and is naturally identical for both polarizations. The JPR resonance occurs in the optimally-doped sample at 250 $cm^{-1}$ (31 meV) and at 60 $cm^{-1}$ (7.5 meV) with the $y$ = 6.67 UD sample. After conversion of $R(\omega)$ to $\sigma(\omega)$ by Kramers-Kronig methods it immediately becomes evident that the loss of spectral weight below $T_c$ stretches not to $2\Delta$ but through to $4\Delta$. This signals that the excitation edge is to be associated here not with pair breaking and the liberation of two unpaired carriers to $E_F$, but with the raising of a bound pair as a unit within the condensate into the upper Bogliubov hole band, an excitation of $4\Delta$ per pair. Above this energy a very interesting hump is recorded. Below $T \approx T_c$, whether resulting from field or temperature change, there is manifest in the spectra a clear isosbestic point at energy $4\Delta$ (see figure 4) with the transfer of spectral weight from the region between $3\Delta$ and $4\Delta$ *up* into the range between $4\Delta$ and $5\Delta$. In $YBCO_{6.67}$ the isosbestic $4\Delta$ point sits at 50 meV (400 $cm^{-1}$) while in $YBCO_{6.95}$ it sits at 80 meV (640 $cm^{-1}$). The hump developing between $4\Delta$ and $5\Delta$ has the appearance of a pseudo-phonon and clearly is related to the $B_{1g}$ feature in the Raman spectra at this energy (fig.2), ascribed above to local pairs, now excited free of the influence of the condensate to somewhat above $E_F$. In the optimally doped sample the spectral region indeed extends down to somewhat lower energy and becomes also suitably lowered in intensity (see [47], figs. 5, 6 and 8].

Let us now look at some of the changes specifically wrought in these spectra by application of the magnetic fields. In underdoped material there is in evidence a marked shortfall, as noted above, between $\rho_s$ as it is directly measured and the integrated spectral weight in $\sigma_1(\omega)$ as generated here when imposing integration cut-offs $\Omega_c \leq 10\Delta$ (~ 150 meV or 1200 $cm^{-1}$). Even for the relatively small fields used, $\rho_s(H)$ is uniformly observed to fall away very appreciably with $|H|$, in a manner akin to thermal excitation. There is, though, pointed distinction here with what is evaluated for $\Delta W(\Omega_c,H)$. $\Delta W$ under $H//c$ moves to meet $\rho_s$ quite quickly, but for $H\perp c$ it progressively decreases retaining its separation from $\rho_s$. This implies that high-energy contribution to the spectral weight is not so damaged with the latter field orientation, for which the flux vortices do not lead to superconducting phase disorder between layers. These high-energy contributions to $\Delta W$ and $\rho_s$ prove widely forthcoming to energies much greater than the upper $10\Delta$ cut-off imposed here. Locally around the isosbestic point, $H\perp c$ has a more marked effect upon spectral weight changes than does $H//c$ (fig.9). This signals that here there arises more simply than loss of phase coherence: there occurs definite extensive transfer of optical spectral weight back to higher energy. $H\perp c$ likewise has a more striking effect upon the JPR feature than does $H//c$; the latter simply depresses $\rho_s$ and $\omega_{JPR}$ somewhat, but $H\perp c$ induces additional new resonances.

The observation that the primary effect of a magnetic field on superconductivity is to break phase coherence rather than actually to break pairs explains directly why all the above effects observed in these remarkably small fields are found to cease near $T_c$. Above $T_c$ one still has the local pairs but these lack phase coherence. The fact that this general pattern is somewhat altered for the changes near the isosbestic point registers that those changes are intimately connected with the excitation itself, the local pairs being in the course of transfer out of the body of the coherent condensate. The ***E***//*c* experiments accordingly have yielded invaluable information about the nature and sustenance of the global condensate.

## §5. Conclusions.

Very recent Raman and optical data dealing with optimal and somewhat hole-rich samples for a variety of HTSC systems have been closely examined from the negative-*U*, boson-fermion crossover point of view. These results all find ready accommodation within that framework; notably the contrasting development of $A_{1g}$, $B_{1g}$ and $B_{2g}$ polarized electronic Raman spectra with $p$ and $T$, and the isosbestic nature of the thermal modification to the infrared optical conductivity occurring about energy $4\Delta$. Both the Raman and IR spectra bear clear imprint of the changes due to quasi-particle incoherence setting in at $p = 0.185$, and bringing about marked changes in electrical transport behaviour, presently a renewed centre of interest. The origin of the incoherence has been presented as resulting from *e*-on-*e* to *b* resonant charged boson production and the ensuing *e*-on-*b* scattering, the latter dominant as the boson population sustained maximizes. Maximum pair condensation energy occurs precisely as quasiparticle incoherence under this carrier scattering is reached at $p = 0.185$. There metallic screening is at a minimum, permitting the establishment of the highest net pair population of local (negative-*U*) and induced carrier pairings. Magnetic field effects on the level of superconductive coherency obtaining prove highly illuminating.

## Figure Captions

**Figure 1.** Slightly modified scale diagram taken from [21], indicating relative and absolute energies across the mixed-valent range from $Cu_{II}$ to $Cu_{III}$ of (i) the bottom of the $\sigma^{*}pd_{x2-y2}$ band, (ii) the Fermi level, and (iii) the local pair fluctuational state $^{10}Cu_{III}^{2-}$. This state is shown (i) optimally resonant with $E_F$ at $p$ = 0.16, (ii) bound relative to $E_F(p)$ at $p$ = 0 with antinodal binding energy $U$ = 75 meV, and (iii) fully unbound by $p$ = 0.28 relative to $E_F$ at that stage. The width assigned to the local pair state derives from the intrinsic disorder of the doping. For both limiting concentrations of $p$ = 0 and 1.0 the system is Mott insulating. $2\Delta_o(p)$ relating to the induced superconducting condensation is represented by the vertical hatching, and maximizes for $p = {}^{1}/_{6}$ at around 40 meV in YBCO. For a representation of how $U$ changes with **k** vector around the Fermi surface both as a function of $p$ and of the system covalence see figure 4 in [17].

**Figure 2.** Development of figure 3 from [35] by Masui *et al*, indicating the evolution of the location of the low temperatures peaks in the $A_{1g}$ and $B_{1g}$ electronic Raman spectra across a range of $p$ values from optimally doped to quite highly overdoped samples of $(Y_{1-x}Ca_x)Ba_2Cu_3O_{7-\delta}$. The $p$ values have here been generated using various combinations of $x$ and $\delta$, and the final behaviour shown to depend solely on $p$ itself, not on $x$ or $\delta$ individually. The $A_{1g}$ and $B_{1g}$ spectra come together above the critical value $p_c$ = 0.185 (see [14], [23] and [24] for significance of this doping level). Below $p_c$ the $B_{1g}$ spectra, which track the local pair binding energy, display a strong increase in Raman scattering peak excitation energy towards lower $p$, but it is the $A_{1g}$ peak that grows there in relative intensity, as the electronic screening diminishes with increasing loss of quasiparticle coherence below $p_c$ under the chronic *e*/*b* scattering. Note the high level of noise on the primary data of figure 2 in [35], and the fact that the phonon spectra have to be subtracted out, make precise identification of these electronic Raman peak energies difficult to specify to better than 40 $cm^{-1}$ (or 5 meV).

**Figure 3.** Amplification of figure 1 in [40c] from von Heumann *et al*, dealing with changes in spectral weight on cooling through $T_c$, as deduced from the low energy optical conductivity spectra for materials drawn from a variety of HTSC systems and distributed across a wide range of $p$ value. As in figure 2, the critical $p_c$ value of 0.18(5) manifests itself in striking fashion, being here where the low-energy spectral weight change $\Delta W$ (cooling through $T_c$) very rapidly, if not discontinuously, changes sign. $\Delta W$ is here evaluated from the partial sum-rule rule due to Ferrell, Glover and Tinkham connecting the amplitude of the superconducting condensate (at $\omega$ = 0), the superfluid density $\rho_s$, to the integrated changes in $\sigma_1(\omega)$ across $T_c$

$$\rho_{s,r} = \int_{0+}^{\Omega} d\omega \,.\{\sigma_{1,r}^{N}(\omega) - \sigma_{1,r}^{sc}(\omega)\} \;.$$

For classical superconductors the cut-off frequency $\Omega$ is typically only ~ $5\Delta$, and in a one-band tight-binding situation $\int\sigma_{1,r}(\omega).d\omega$ is proportional to the electronic kinetic energy decrease. By contrast in the present case a 1 eV cut-off has been imposed. Below $p_c$ the data appear on separate traces dependent on whether for mono-layer, bi-layer or tri-layer material.

**Figure 4.** Real part of the optical conductivity for $YBCO_{6.67}$ and $YBCO_{6.95}$. Attention is drawn to the isosbestic intensity changes with temperature change across energies $4\Delta_o$. These points are marked by vertical arrows. Note the logarithmic frequency scale in the lower panel. Between $3\Delta$ and $4\Delta$ the intensity falls on cooling and is transferred up to the $4\Delta$ to $5\Delta$ range. The figure is based on figure 5 of [47] and is for $H = 0$. (400 $cm^{-1}$ ≡ 50 meV)

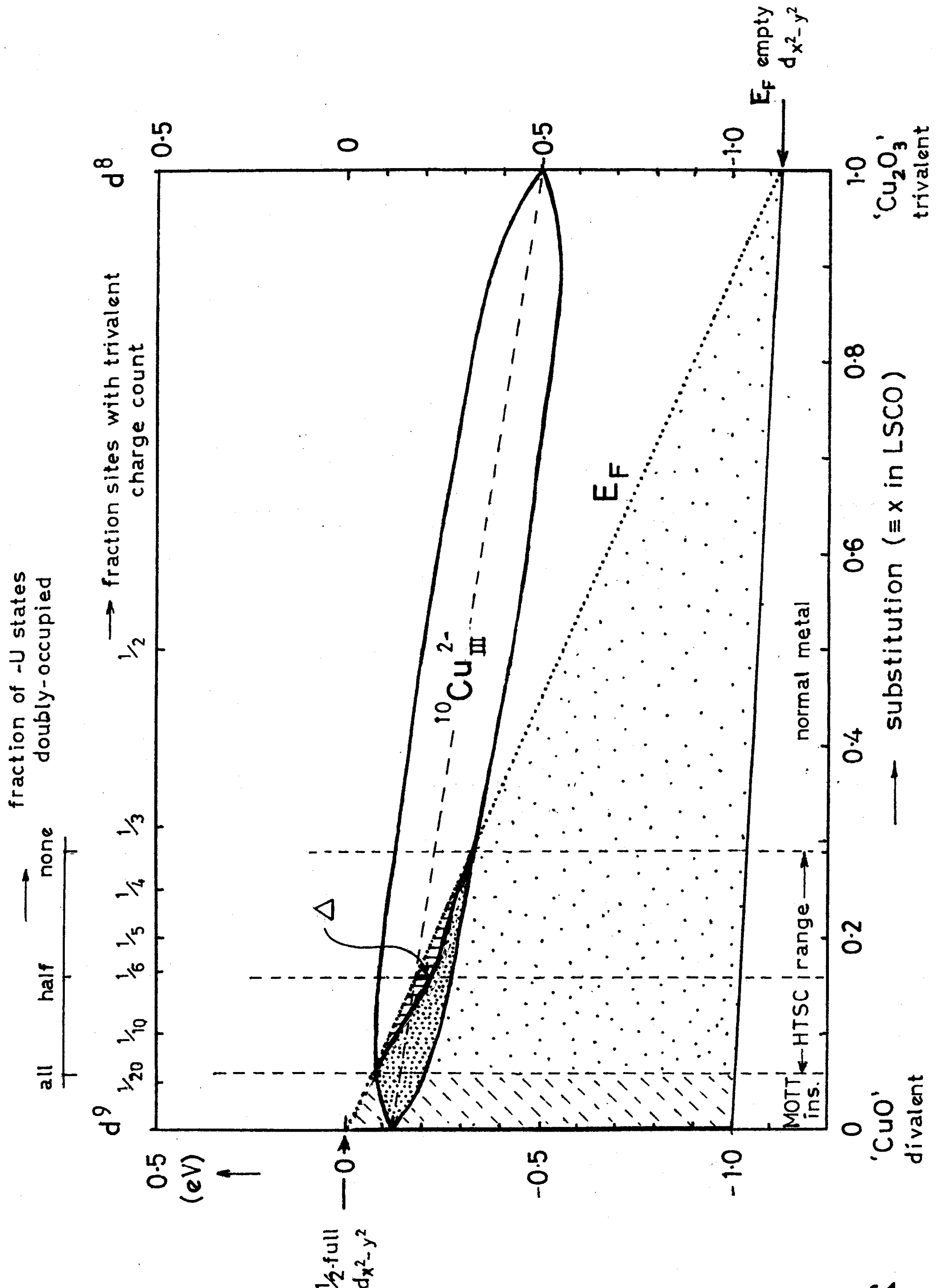

fraction of -U states
doubly-occupied
all
half
none
→
$\frac{1}{20}$ $\frac{1}{10}$ $\frac{1}{6}$ $\frac{1}{5}$ $\frac{1}{4}$ $\frac{1}{3}$ $\frac{1}{2}$
→ fraction sites with trivalent charge count
$d^9$
$d^8$
0·5
(eV)
0
-0·5
-1·0
$E_F$ ½-full $d_{x^2-y^2}$
$E_F$ empty $d_{x^2-y^2}$
$^{10}Cu^{2-}_{III}$
Δ
$E_F$
MOTT ins.
←HTSC range→
normal metal
0
0·2
0·4
0·6
0·8
1·0
'CuO' divalent
'$Cu_2O_3$' trivalent
→ substitution (≡ x in LSCO)

f1

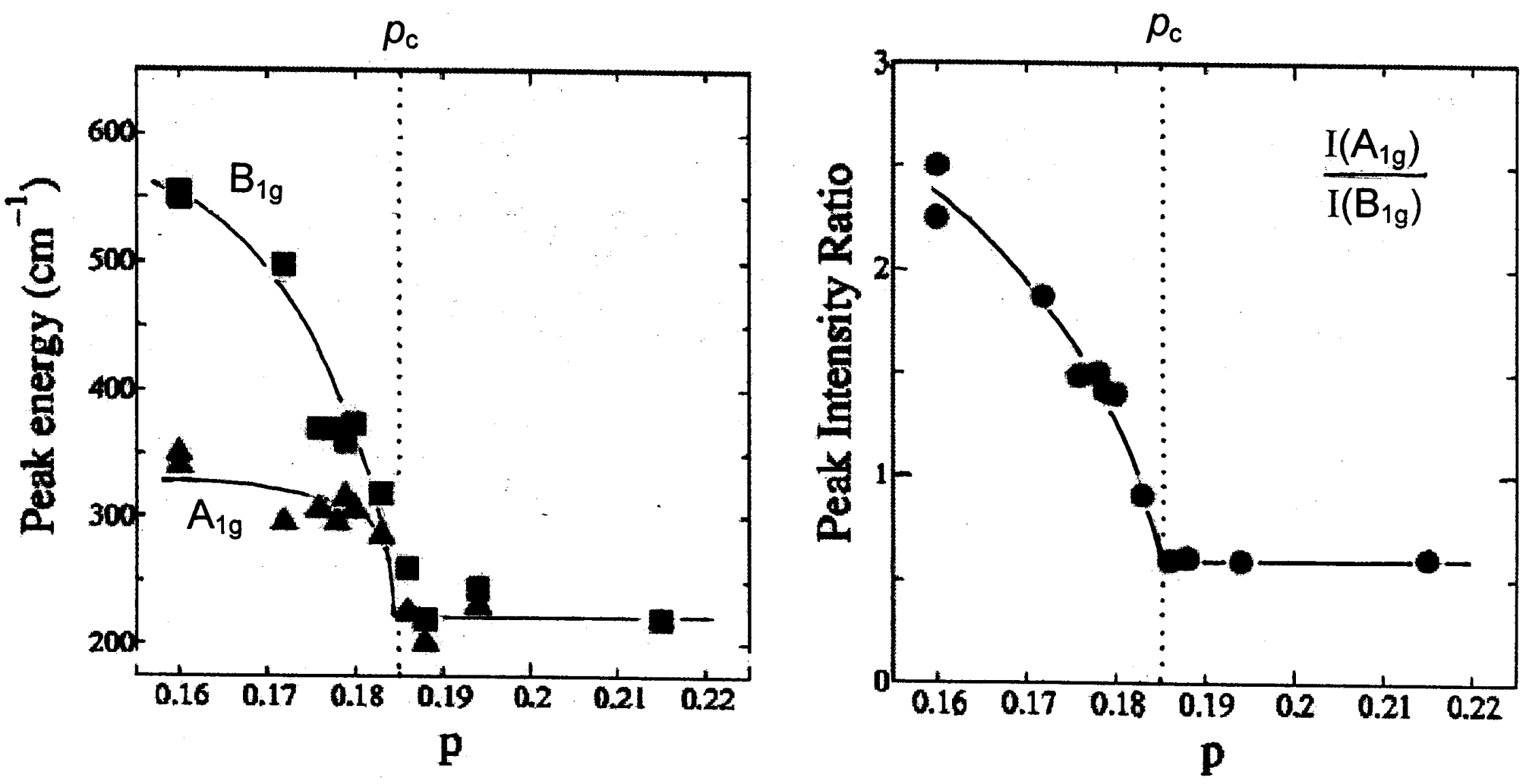

f2

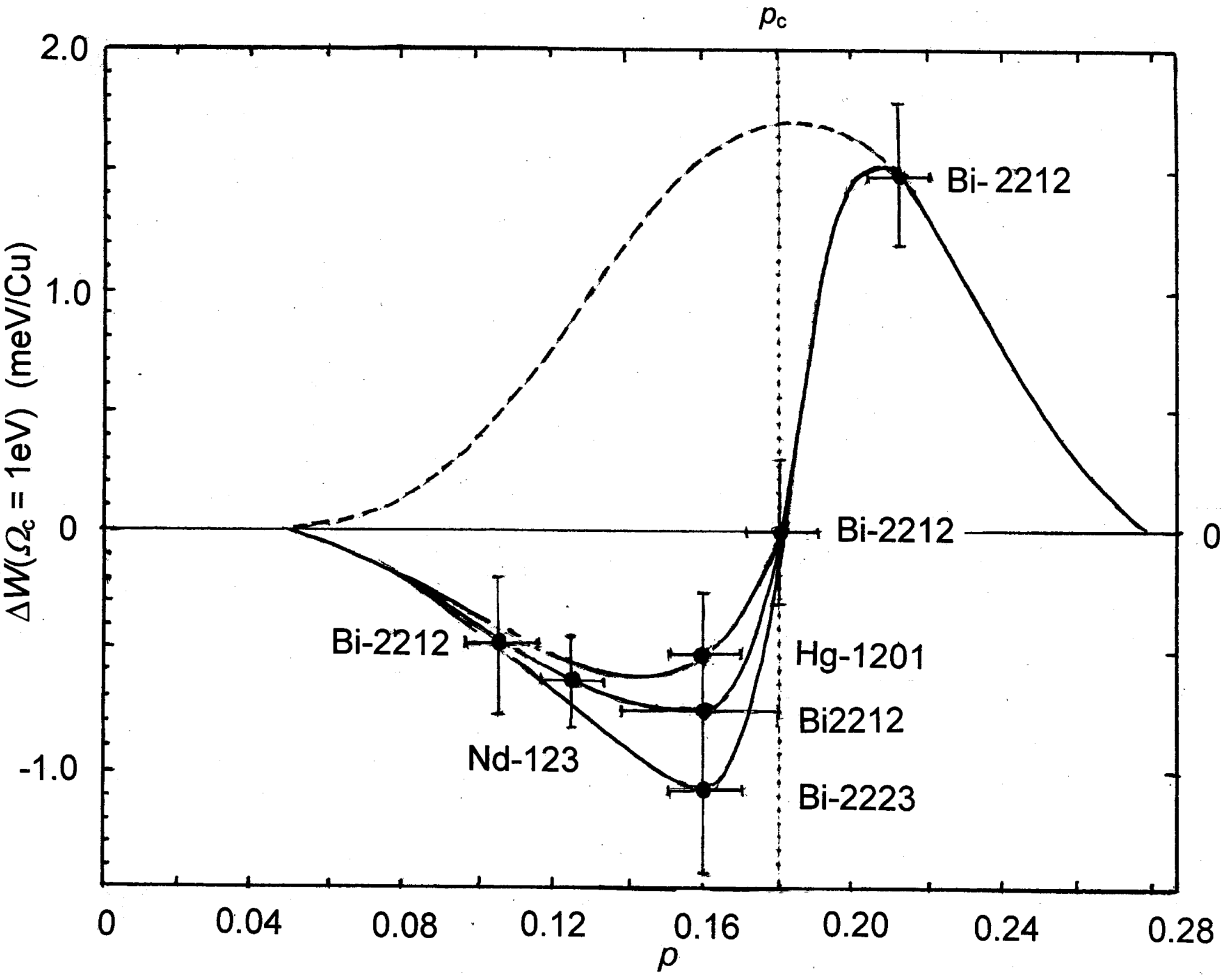

f3

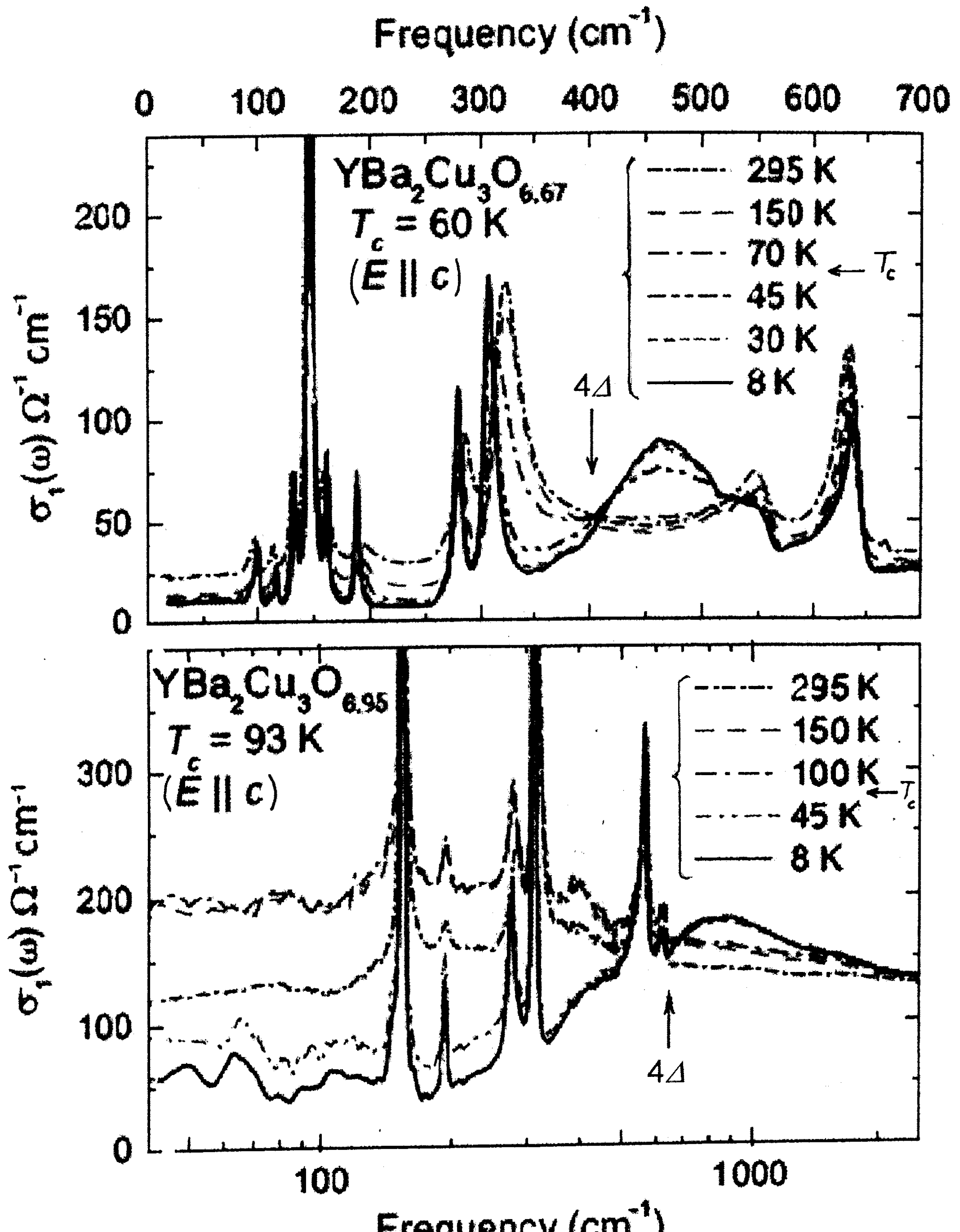

f4